\documentclass[twocolumn,superscriptaddress]{revtex4}
\usepackage{amsmath,amssymb,amsfonts,amsthm,mathtools,bm}
\usepackage{graphicx}
\usepackage[colorlinks=true,
linkcolor=blue,citecolor=blue,urlcolor=blue]{hyperref}
\newcommand\ii{\mathrm{i}}
\newcommand\dd{\mathrm{d}}\newcommand\sech{\mathrm{sech}}
\newcommand\Qw{Q_{\mathrm{w}}}\newcommand\vs{v_{\mathrm{s}}}
\newcommand\vb{v_{\mathrm{b}}}\newcommand\vrr{v_{\mathrm{r}}}
\newcommand\ds{\mathrm{ds}}\newcommand\gs{\mathrm{gs}}
\newcommand\lw{\mathrm{lw}}\newcommand\cp{\mathrm{cp}}
\newcommand\im{\mathrm{im}}\newcommand\bs{\mathrm{bs}}
\begin{document}
\title{Topological charges engineered by density zeros}
\author{Xiao-Lin Li}
\affiliation{School of Physics, Northwest University, Xi'an 710127, China}
\author{Ming Gong}
\affiliation{CAS Key Laboratory of Quantum Information, University of Science and Technology of China, Hefei 230026, China}
\affiliation{Synergetic Innovation Center of Quantum Information and Quantum Physics, University of Science and Technology of China, Hefei 230026, China}
\affiliation{Hefei National Laboratory, University of Science and Technology of China, Hefei 230026, China}
\author{Yu-Hao Wang}
\affiliation{School of Physics, Northwest University, Xi'an 710127, China}
\author{Li-Chen Zhao}
\email{zhaolichen3@nwu.edu.cn}
\affiliation{School of Physics, Northwest University, Xi'an 710127, China}
\affiliation{Shaanxi Key Laboratory for Theoretical Physics Frontiers, Xi'an 710127, China}
\affiliation{NSFC-SPTP Peng Huanwu Center for Fundamental Theory, Xi'an 710127 and Hefei 230026, China}
\date{\today}
\begin{abstract}
We propose an approach to engineer topological charges by manipulating wave function zeros, departing from conventional energy gap tuning. Inspired by Dirac's monopole theory, we demonstrate that the winding number of a toroidal Bose-Einstein condensate can be precisely controlled via density zeros of dark solitons---each zero induces a phase singularity, triggering a quantized change in the winding number, with the direction of zero relative velocity crossing dictating the increment or decrement. This method enables arbitrary integer tuning of the winding number through accelerating multiple solitons or soliton-to-vortex transitions, and is feasible with current experimental parameters for $^{87}$Rb condensates. The explicit relation between winding number and rotation current on the ring could be used to design a compact rotation sensor.  These findings open new avenues for designing topological systems and advancing quantum simulation technologies.
\end{abstract}
\maketitle
Topological charge stands as a cornerstone in modern physics, encapsulating the intrinsic topological properties of field configurations, topological defects, and spacetime geometries \cite{MorandiSpringer1992,ThoulessWorldSci1998,EschrigSpringer2011}. Rooted in the global topological characteristics of physical systems, it serves as a universal descriptor of non-trivial states across condensed matter physics \cite{ChaikinCambridge1995,BhattacharjeeSpringer2017}, topological photonics \cite{OzawaRMP2019,WangFP2022}, and cold atomic physics \cite{CooperRMP2019,ZhangAP2019}. Defined as an integer-valued invariant, topological charge distinguishes topologically distinct states through their homotopy classes, underpins quantized observables, and governs emergent phenomena such as topological protection and quantized transport---exemplified by the quantum Hall effect \cite{HeNSR2014,KlitzingNRP2020,ChangRMP2023} and Thouless pumping \cite{XiaoRMP2010,CitroNRP2023}. Critically, its robustness against local perturbations and symmetry-preserving deformations renders it indispensable for understanding and engineering stable topological phases \cite{HasanRMP2010,QiRMP2011}.

For decades, the manipulation of topological charges has primarily relied on topological band theory \cite{HasanRMP2010,QiRMP2011,BansilRMP2016}, where topological phase transitions are induced by closing and reopening energy gaps between occupied and unoccupied bands. While powerful, this paradigm is constrained by the necessity of energy gap modulation, limiting flexibility in engineering topological states. In contrast, Dirac's seminal work on magnetic monopoles \cite{DiracPRSLA1931} hinted at a profound link between wave function zeros and topological charges---a connection largely untapped for active control of topological properties \cite{NyePRSLA1974,DuanPRE1999,HePRE2021}. Recent studies \cite{RayNature2014,RayScience2015,ZhaoPRE2021,YuPRA2024} have further revealed that wave function zeros in real space carry non-zero topological charges, suggesting their potential as a novel knob for engineering topology.

Here, we report a paradigm-shifting approach to engineer topological charges by harnessing the zeros of wave functions, departing from conventional energy gap tuning. We demonstrate that the winding number---a key topological charge---of a toroidal Bose-Einstein condensate (BEC) can be precisely controlled through the density zeros of dark solitons, with analogous behavior observed in two-dimensional systems involving vortices. The core mechanism underlying this control is unveiled: each zero of the wave function induces a phase singularity, which universally triggers a quantized change in the winding number of the toroidal condensate.
By driving solitons with external forces, we show that repeated crossings of zero relative velocity lead to stepwise adjustments of the winding number---enabling not only binary changes but also arbitrary integer tuning when leveraging periodic oscillations of spin solitons.  These results provide a new avenue for designing topological systems, offering unprecedented flexibility in manipulating topological charges without relying on energy gap modulation.

We consider a binary toroidal BEC as a general platform to demonstrate the protocols for the topological charges engineered by density zeros. Rescaling the atomic mass and Planck's constant to be $1$, the dimensionless mean-field energy for the system in polar coordinates is expressed as \cite{DalfovoRMP1999,PitaevskiiOUPress2016}
\begin{align}
E=\int\bigg[
&\frac12\big(|\nabla\psi_\ds|^2+|\nabla\psi_\lw|^2\big)
-FR\theta|\psi_\lw|^2
\nonumber\\
+&\frac12\omega^2\big(r-R\big)^2\big(|\psi_\ds|^2+|\psi_\lw|^2\big)
\nonumber\\
+&\frac{g_\ds}{2}|\psi_\ds|^4+\frac{g_\lw}{2}|\psi_\lw|^4
+g_\cp|\psi_\ds|^2|\psi_\lw|^2\bigg]r\dd r\dd \theta,
\end{align}
where $\psi_\ds$ and $\psi_\lw$ represent the dark solitons and localized waves (e.g., impurities or bright solitons), respectively. These localized waves are trapped by dark solitons via the nonlinear interaction $g_\cp|\psi_\ds|^2|\psi_\lw|^2$. The dimensionless coupling constants $g_\ds$, $g_\lw$ and $g_\cp$ characterize the corresponding interaction strengths among atoms. To avoid the snaking instabilities
\cite{CarreteroNonlinearity2008,FrantzeskakisJPA2010}, the trapping frequency $\omega$ must be sufficiently large in the radial coordinate $r$, making the condensate appear as a quasi-one-dimensional ring with radius $R$ along the angular direction $\theta$. The external forces $F$ are applied to accelerate dark solitons without perturbing their background densities, providing a practical way for controlling their motions \cite{ZhaoPRA2020,MengJPB2024}.

\begin{figure}[t]
\includegraphics[width=8.5cm]{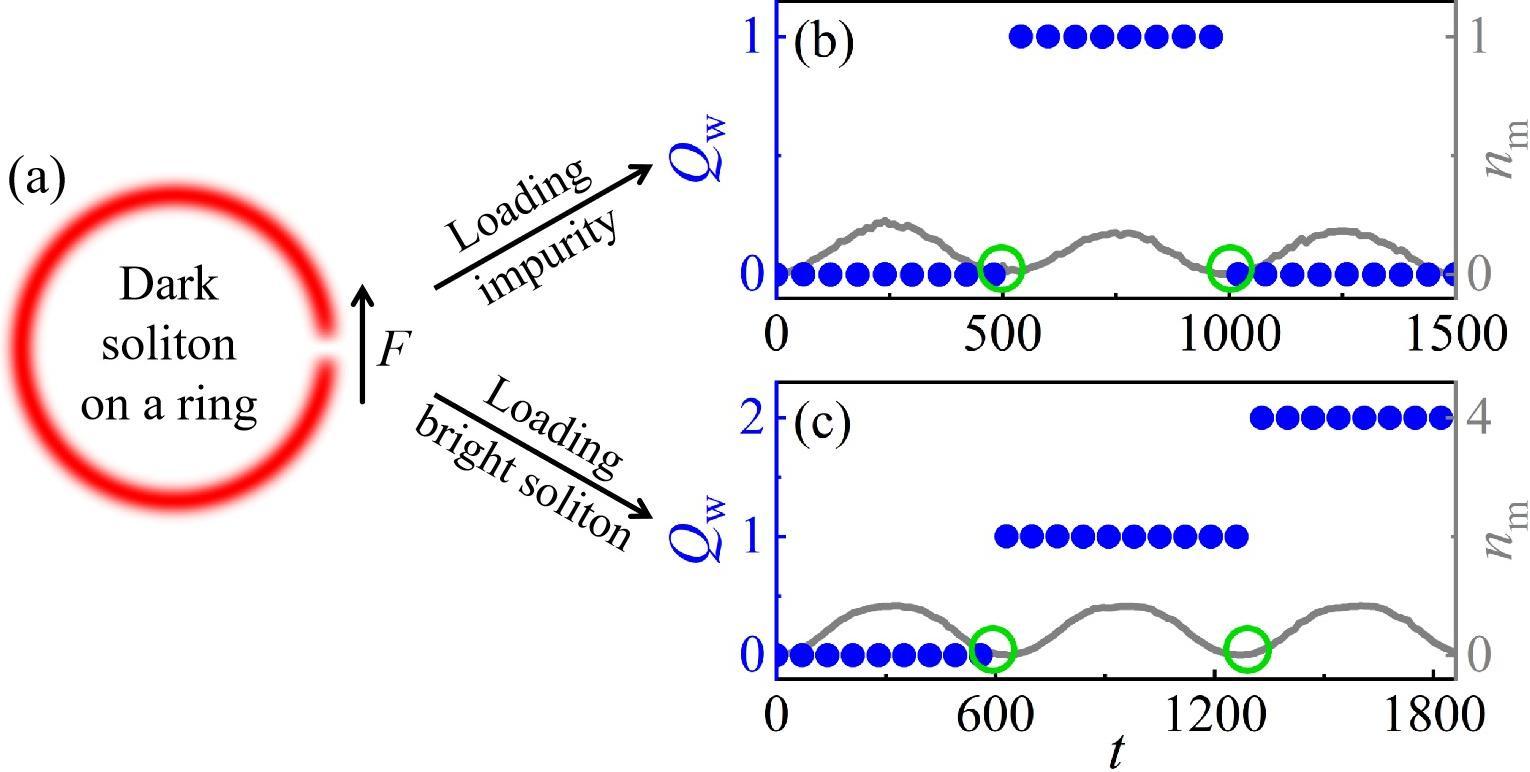}
\caption{\label{Fig1}(a) Schematic representation of one dark soliton driven by a force $F$ in a toroidal Bose condensate, and corresponding evolution of winding number $\Qw$ for loading impurity in (b) or bright soliton in (c). Gray lines and green circles denote evolution of soliton's density minimum $n_{\mathrm{m}}$ and occurrence of density zero, respectively. Employing cosinusoidal force as $F=-0.05\cos(2\pi t/1000)$ in (b) and constant force $F=0.005$ in (c), and setting parameters as $R=50$, $|\psi_{\gs}|^2=1$, $v_{\mathrm{r, s, b}}|_{t=0}\approx0$, $\omega=1$, $g_{\ds}=g_{\cp}=1$, $g_{\im}=0$, $\varepsilon=1/\sqrt{10}$ in (b) and $\omega=3$, $g_{\ds}=3$, $g_{\cp}=2$, $g_{\bs}=1$ in (c), respectively.}
\end{figure}

We first consider a dark soliton coupled to an impurity, where the initial dark soliton can be expressed as
\begin{align}
\psi_\ds=\bigg[
\ii\vrr+\sqrt{1-\vrr^2}\tanh\big(\sqrt{1-\vrr^2}R\theta\big)\bigg]
e^{i\vb R\theta}\psi_\gs,
\end{align}
and the impurity as $\psi_\im=\varepsilon\sech
\big[\sqrt{1-\vrr^2}R\theta\big]e^{i\vs R\theta}\psi_\gs$ with replacing the localized wave $\psi_\lw$ with $\psi_\im$. The toroidal ground state $\psi_\gs\equiv1$ is obtained by the imaginary-time evolution method \cite{YangSIAM2010,BaoKRM2013}. The parameters $\vs$, $\vb$ and $\vrr\equiv\vs-\vb$ represent the soliton velocity, background density velocity, and relative velocity between them. Furthermore, the single-valued nature of the order parameter requires that the phase variation around the ring satisfies \cite{BrandJPB2001,MateoPRA2015,ShamailovPRA2019}
\begin{align}\label{Qw}
\Delta\phi+2\pi R\vb=2\pi\Qw,
\end{align}
where $\Delta\phi$ and $2\pi R\vb$ account for the phase shifts induced by the dark soliton and background current, respectively. The integer $\Qw$ represents the winding number, defined as the number of of times the phase winds in a closed loop counterclockwise around the ring. The explicit relation between winding number and the rotation current could be used to design a rotation sensor, since it is easier to measure the quantized winding number and the phase jump of soliton in experiments.  Notably, the soliton's phase jump $\Delta\phi$ is related to its relative velocity $\vrr$, which provides the the possibilities for engineering winding number $\Qw$.
\begin{figure}[t]
\includegraphics[width=8.5cm]{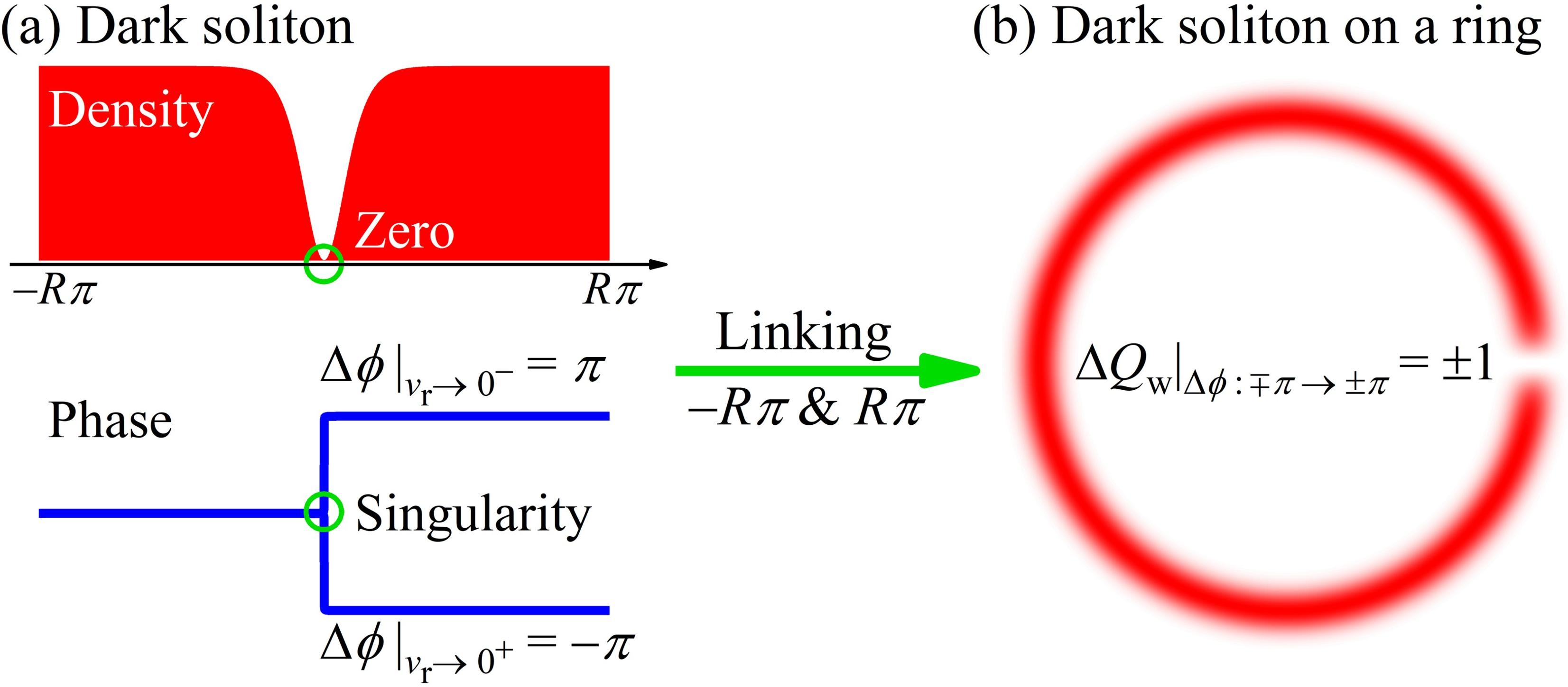}
\caption{\label{Fig2}Schematic of engineering winding number $\Qw$ in a toroidal condensate via density zero of dark soliton. For soliton in (a), emergence of density zero, where relative velocity cross zero, induces a total phase jump of $\pm2\pi$. Linking ends of soliton to form the ring geometry in (b), this $\pm2\pi$ phase jump yields a change of $\pm1$ in winding number $\Qw$.}
\end{figure}

\begin{figure*}[tbp]
\includegraphics[width=17cm]{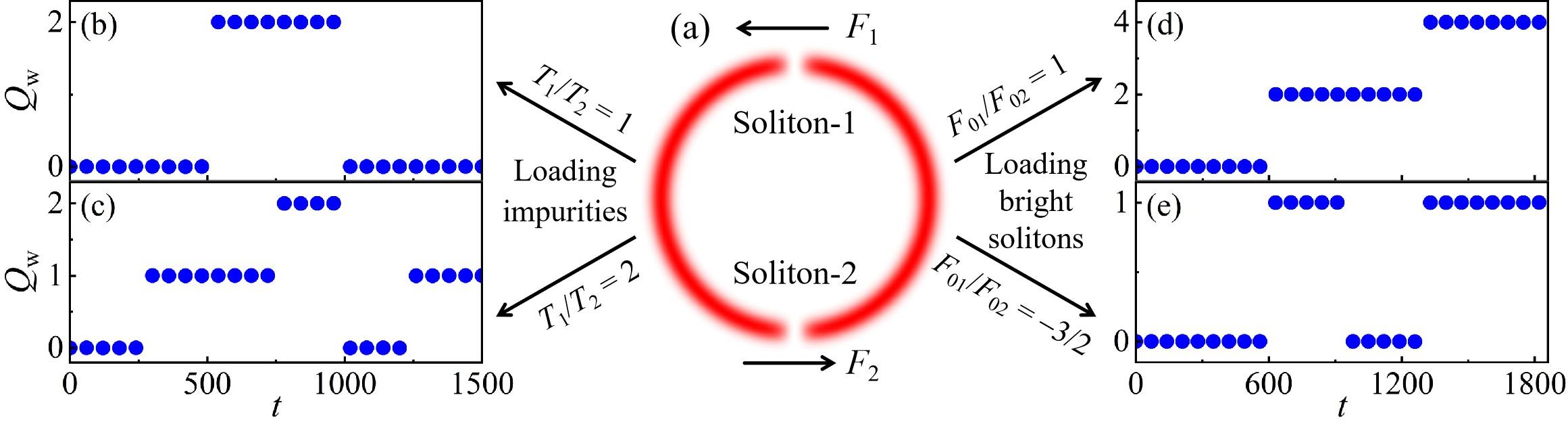}
\caption{\label{Fig3}(a) Schematic of two well-separated dark solitons driven by forces $F_j$ in a toroidal Bose condensate, and corresponding evolution of winding number $\Qw$ for loading impurities in (b)-(c) or bright solitons in (d)-(e). Employing cosinusoidal driving force as $F_j=-0.05\cos(2\pi t/T_j)$ with $T_1=1000$ in (b)-(c), and constant driving force as $F_j=F_{0j}$ with $F_{01}=0.0005$ in (d)-(e), and setting parameters as $R=150$, $|\psi_{\gs}|^2=1$, $v_{\mathrm{r, s, b}j}|_{t=0}\approx0$, $\omega=1$, $g_{\ds}=g_{\cp}=1$, $g_{\im}=0$, $\varepsilon_j=1/\sqrt{10}$ in (b)-(c) and $\omega=3$, $g_{\ds}=3$, $g_{\cp}=2$, $g_{\bs}=1$ in (d)-(e), respectively.}
\end{figure*}

Employing a cosinusoidal force $F=-0.05\cos(2\pi t/1000)$ (as shown in Fig. \ref{Fig1} (a)), the dynamics of dark soliton are simulated using the Fourier pseudo-spectral method with Strang splitting \cite{YangSIAM2010,BaoKRM2013}. Due to its negative effective mass, the dark soliton accelerates opposite to the force. Initially, the soliton moves counterclockwise; upon the reversal of force, its relative velocity $\vrr$ asymptotically approaches zero, subsequently reversing to induce clockwise motion; and vice versa. This process repeats periodically with each force reversal, as detailed in the full video of Supplemental Material \cite{SM}.

The evolution of winding number $\Qw$ and soliton density minimum are shown in Fig. \ref{Fig1} (b). It is seen that the density zeros (occur at $\vrr=0$ moments) precisely correspond to the abrupt jump points of $\Qw$. Moreover, the crossing direction of the zero relative velocity determines whether the winding number $\Qw$ increases or decreases. In recent experiments
\cite{RamanathanPRL2011,WrightPRL2013,RyuPRL2013,EckelNature2014}, the winding number can only be manipulated by a rotating weak link in the absence of dark solitons, which is achieved by changing the quantized persistent currents. Here we show that a much easier way to achieve this is by changing $\vrr$ and the soliton motion can also modulate the continuous persistent currents.

Notably, the above protocol can only realize two states, characterized by the winding number of $\Qw|_{t=0}\pm1$. This limitation prompts us to consider a more general question: whether we can manipulate the winding number to an arbitrary integer number. In the following, we propose a feasible protocol to successively increase or decrease winding number based on the periodic oscillation of the spin soliton driven by a constant force $F$ \cite{ZhaoPRA2020,MengPRA2022}. Here, we replace the localized wave $\psi_\lw$ with the bright soliton $\psi_\bs=\sqrt{1-\vrr^2}\sech\big(\sqrt{1-\vrr^2}R\theta\big)
e^{i\vs R\theta}\psi_\gs$ and set $g_\ds+g_\bs=2g_\cp$. The basic characteristic of spin soliton admits a uniform total density $|\psi_\ds|^2+|\psi_\bs|^2=|\psi_\gs|^2$, in contrast to the total density admits a dip for previous dark-bright solitons \cite{BuschPRL2001,BeckerNP2008}.

Employing constant force $F=0.005$ on bright soliton, the spin soliton exhibits a periodic oscillation due to the negative-positive mass transition \cite{ZhaoPRA2020,MengPRA2022}. Thus the dark soliton will initially move in a clockwise direction and then move in a counterclockwise direction. This process will also be repeated accordingly, seeing the full video of dynamics in the Supplemental Material \cite{SM}. Notably, the density zero only emerges when the relative velocity $\vrr$ crosses zero in the same manner, where the spin soliton admits the negative mass. Therefore, a step-like increment in the winding number $\Qw$, as presented in Fig. \ref{Fig1} (c). However, it should be emphasized that the soliton eventually breakdown when the winding number $Q_{\rm w}$ is too large, partly due to that there are sound waves or dispersive waves emitted during each oscillation process. For the parameters used in Fig. \ref{Fig1} (c), the maximum observed value for $Q_{\rm w}$ is approximately $30$. A larger maximum value of $Q_{\rm w}$ can be obtained with weaker constant force, such as the maximum observed value for $Q_{\rm w}$ is more than $250$ when constant force $F=0.0025$.

To understand this change of topological charges accompanying with the density zero, we begin to investigate the correlation between density zero and phase jump of dark soliton. In Refs. \cite{ZhaoPRE2021,YuPRA2024}, we have shown that the wave function possesses a phase jump of $\pm\pi$ at the density zero, with the sign determined by the approaching direction. For one dark soliton, we note that
\begin{align}
\lim_{\vrr\rightarrow0^{\mp}}\Delta\phi=\pm\pi.
\end{align}
Therefore, as the zero relative velocity is approached from different directions ($+$ or $-$), the phase jump of $\pm\pi$ direction varies. For instance, when $\vrr$ changes from a positive value to zero, a phase jump of $-\pi$ occurs; and when $\vrr$ changes from zero to a negative value, a phase jump of $\pi$ occurs. Then a total phase jump of $2\pi$ occurs, as demonstrated in Fig. \ref{Fig2} (a); and vice versa. In one-dimensional settings with open boundaries, this $2\pi$ phase jump occurs in accelerating dark soliton, yet it is not observable. Linking the two ends of a one-dimensional dark soliton to form a ring configuration, we uncover a significant finding this phase jump will become observable, which is the essential idea used in geometric phase. As shown in Eq. (\ref{Qw}), the $\pm2\pi$ phase jump induces a change of $\pm 1$ in the winding number $\Qw$, as demonstrated in Fig. \ref{Fig2} (b). Explicitly, each crossing of zero relative velocity induces a change in the winding number, with the direction of the crossing determining whether the winding number increases or decreases.

We next investigate two well separated dark solitons with coupled impurity atoms driven by two weak forces $F_{1,2}$, as shown in Fig. \ref{Fig3} (a), which can be used as a platform to demonstrate the arbitrary tunability of winding number. The two dark solitons are described as
$\psi_\ds=
\big[\ii\vrr+\sqrt{1-\vrr^2}
\tanh\big(\sqrt{1-\vrr^2}R\theta_1\big)\big]
\big[\ii\vrr+\sqrt{1-\vrr^2}
\tanh\big(\sqrt{1-\vrr^2}R\theta_2\big)\big]
e^{\ii\vb R\theta}\psi_\gs$
and coupled impurities as $\psi_\im=
\sqrt{1-\vrr^2}\big[\sech\big(\sqrt{1-\vrr^2}R\theta_1\big)
e^{\ii\vs R\theta_1}
+\sech\big(\sqrt{c^2-v_{\rm r}^2}R\theta_2\big)
e^{\ii\vs R\theta_2}\big]\psi_{\rm gs}$ or bright solitons as
$\psi_\im=
\varepsilon\big[\sech\big(\sqrt{1-\vrr^2}R\theta_1\big)
e^{\ii\vs R\theta_1}
+\sech\big(\sqrt{c^2-v_{\rm r}^2}R\theta_2\big)
e^{\ii\vs R\theta_2}\big]\psi_{\rm gs}$ with $\theta_{1,2}=\theta\mp\pi/2$. Each soliton is driven by a weak force $F_{1,2}$. We present different evolution of winding number by adjusting period ratio $T_1/T_2$ in Fig. \ref{Fig3} (b)-(c) or strength ratio $F_{01}/F_{02}$ in Fig. \ref{Fig3} (d)-(e). When $T_1/T_2=1$ or $F_{01}/F_{02}=1$, the winding number $\Qw$ jumps twice the value observed for one dark soliton (see Fig. \ref{Fig1} (b) and (c)); and when $T_1/T_2\neq1$ or $F_{01}/F_{02}\neq1$, the winding number $\Qw$ evolves in a near random way due to the different frequencies and directions of the density zero crossings, which provides diverse techniques for tailoring topological charges.
\begin{figure}[tbp]
\includegraphics[width=6.5cm]{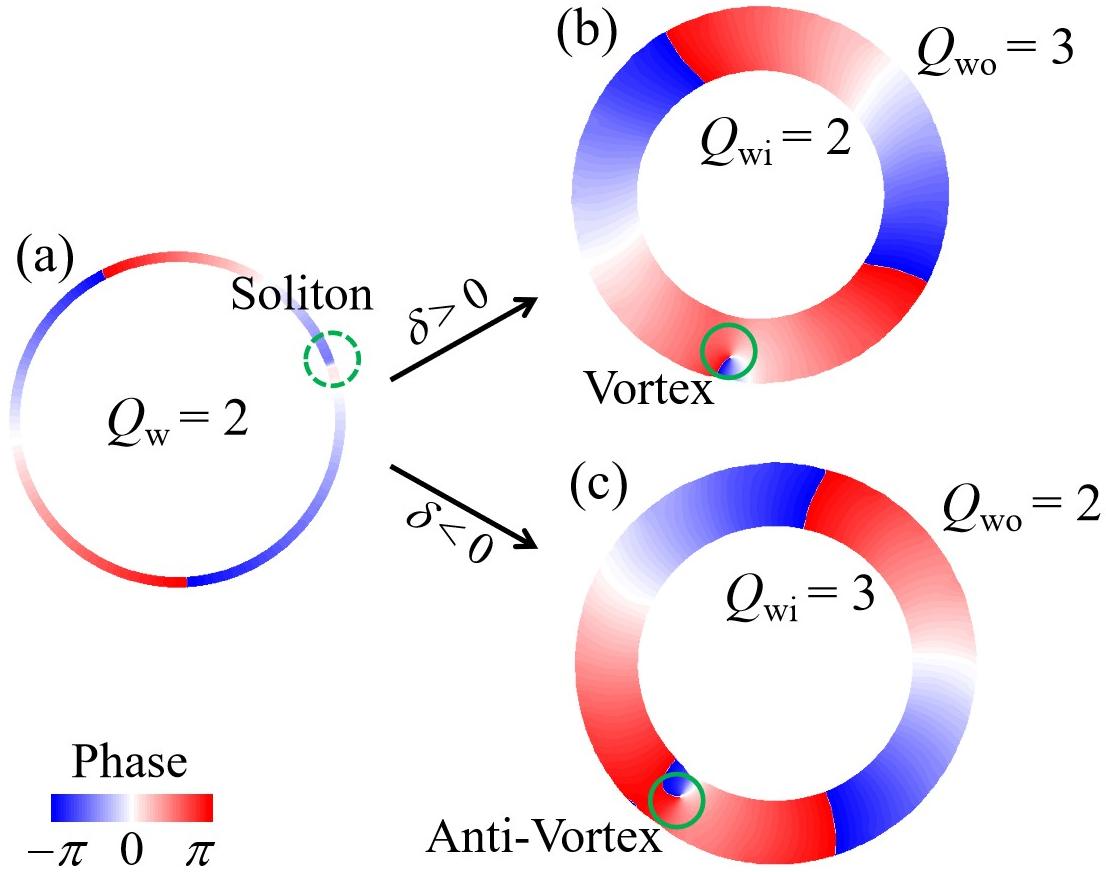}
\caption{\label{Fig4}Engineer of winding number $\Qw$ via quasi-adiabatic transition from a dark soliton to a vortex. (a) Dark soliton in Fig. \ref{Fig1} (c) at $t=1800$ with winding number $\Qw=2$. (b)-(c) Vortex controllably generated by weak asymmetric operation. Employing quasi-adiabatic operation as $\omega\rightarrow\omega-\alpha t$ and introducing weak asymmetric operation as $\delta\int(r-R)|\psi_{\ds}|^2r\dd r\dd\theta$, and setting parameters as $\alpha=-0.00025$, $|\delta|=0.02$, respectively.}
\end{figure}

Vortices, as prototypical topological defects in high-dimensional systems, also exhibit characteristic density zeros whose topological charges fundamentally govern system topology \cite{PitaevskiiOUPress2016,FetterRMP2009}. Engineering these density zeros offers a viable approach to manipulating vortex charges \cite{RichaudPRA2023}. Here, we employ quasi-adiabatic operation $\omega\rightarrow\omega-\alpha t$ to transform dark solitons into vortices. Upon introducing a weak asymmetric operation written as \(\delta\int(r-R)|\psi_{\ds}|^2r\dd r\dd\theta\) in the mean-field energy, the soliton undergoes deterministic evolution into a vortex \cite{LiPRA2024}. As illustrated in Fig. \ref{Fig4}, the generated vortex or anti-vortex independently modulates the winding numbers of the inner (\(Q_{wi}\)) and outer rings (\(Q_{wo}\)) of the toroidal condensate. By combining quasi-adiabatic operations with dark soliton driving processes, \(Q_{wi}\) and \(Q_{wo}\) can be precisely engineered. These results open up additional avenues for topological engineering in high-dimensional scenarios.

We discuss the possibilities to observe the winding number manipulations in real experiments. Let's consider a $^{87}$Rb BEC with two internal states \cite{HallPRL1998,MertesPRL2007,BeckerNP2008} to confirm the results presented in Fig. \ref{Fig1} (b). The related scattering lengths are about $100a_{\rm B}$ ($a_{\rm B}$ being the Bohr radius) \cite{HallPRL1998,MertesPRL2007}. The condensate can be loaded in a toroidal optical dipole trap \cite{RamanathanPRL2011,WrightPRL2013,RyuPRL2013,EckelNature2014} with radial and vertical trapping frequencies of $2\pi\times(100,200){\ \mathrm{Hz}}$ and a radius of $54{\ \mathrm{\mu m}}$. The initial states can be prepared  by imprinting a dark soliton with $2\times10^4$ atoms (the background density of $1.3\times10^{13}{\ \mathrm{cm^{-3}}}$) in state $\vert F=2, m_f=0\rangle$ and filling the density dip with $13$ atoms in state $\vert F=1, m_f=-1\rangle$ \cite{FrantzeskakisJPA2010,BeckerNP2008}. The effective radial length is $1.41\ \xi$ with healing length of $\xi\approx0.76{\ \mathrm{\mu m}}$. A cosinusoidal modulating gradient magnetic field \cite{RayNature2014,RayScience2015,LuoSR2016} with weak peak gradient of $0.07{\ \mathrm{G\cdot cm}^{-1}}$ can drive atomic motion along the angle coordinate, and one jump in the winding number can be observed in about $0.8{\ \mathrm{s}}$. The predictions in this work could be realized through the state-of-the-art experimental techniques.

To conclude, we demonstrate that the topological charge of a toroidal BEC can be well manipulated by engineering its zeros of the wave functions. When the relative velocities between the dark solitons and their background density currents are zeros, the density zeros can occur. We uncover that the unobservable phase jumps of dark solitons crossing density zeros in one-dimensional settings with open boundaries become observable by inducing winding number variations in a ring condensate.  Furthermore, we demonstrate that density zeros in higher-dimensional topological defects such as vortices also enable controlled engineering of topological charges.

Notably, these engineering protocols are feasible with current experimental parameters, as shown by simulations using realistic conditions for $^{87}$Rb
BECs in toroidal traps. The relation between winding number and the rotation current could be used to design a compact rotation sensor.  This work thus opens a new avenue for designing topological systems, offering unprecedented flexibility in manipulating topological charges without relying on energy gap modulation. Such advances may inspire innovative strategies for quantum simulation platforms, where precise control over topological properties is pivotal for exploring exotic quantum phenomena and developing robust quantum technologies \cite{NayakRMP2008,BlochNP2012,TerhalRMP2015,AmicoRMP2022,PoloQT2024}.

\emph{Note added.} Recently, we became aware of a recent experimental work \cite{Rabec2025}, in which the winding number variations were
also observed on a ring BEC and the underlying mechanism is similar to the above one.
\

\

\

\noindent$\displaystyle\textbf{Acknowledgments}$

\noindent$\displaystyle$ L-C Zhao is supported by the National Natural Science Foundation of China (Contracts No. 12375005, No. 12235007 and No. 12247103). M Gong is supported by the Strategic Priority Research Program of the Chinese Academy of Sciences (Grant No. XDB0500000) and the Innovation Program for Quantum Science and Technology (Grants No. 2021ZD0301200 and No. 2021ZD0301500).

\


\end{document}